\documentclass[letterpaper, 10 pt, conference]{ieeeconf}  

\IEEEoverridecommandlockouts

\usepackage{amsmath} 
\usepackage{amssymb}  

\usepackage{amsthm}
\usepackage{booktabs}
\usepackage[ruled,vlined]{algorithm2e}
\SetAlgorithmName{Procedure}{algorithm}{List of Procedures}

\usepackage{tikz}
\usetikzlibrary{shapes,arrows,calc,fit}
\tikzset{
        block/.style = {draw, rectangle,
            minimum height=1cm,
            minimum width=2cm},
        input/.style = {coordinate,node distance=1cm},
        output/.style = {coordinate,node distance=4cm},
        arrow/.style={draw, -latex,node distance=2cm},
        pinstyle/.style = {pin edge={latex-, black,node distance=2cm}},
        sum/.style = {draw, circle, node distance=1cm},
    }
\usepackage{pgfplots} 
\pgfplotsset{compat=newest} 
\pgfplotsset{plot coordinates/math parser=false} 
\newlength\figureheight%
\newlength\figurewidth%
\pgfplotsset{
    every axis plot post/.style={
        line join=round
    }
}

\newlength\defcolwidth%

\usepackage{tikz}
\usepackage{tikzscale}

\definecolor{myorange}{cmyk}{0,0.35,0.85,0} 
\definecolor{mypurple}{cmyk}{0.5,1,0,0} 

\definecolor{matblue1}{rgb}{0,0.4470,0.7410}
\definecolor{matred1}{rgb}{0.85,0.325,0.098}
\definecolor{matyel1}{rgb}{0.9290, 0.6940, 0.1250}
\definecolor{matpur1}{rgb}{0.4940, 0.1840, 0.5560}
\definecolor{matgre1}{rgb}{0.4660, 0.6740, 0.1880}
\definecolor{matblue2}{rgb}{0.3010, 0.7450, 0.9330}
\definecolor{matred2}{rgb}{0.6350, 0.0780, 0.1840}
\definecolor{matgrey1}{rgb}{0.5, 0.6, 0.7}
\definecolor{matpink1}{rgb}{1, 0.07, 0.65}
\definecolor{matblue3}{rgb}{0.07, 0.62, 1}
\definecolor{gray09}{rgb}{0.9, 0.9, 0.9}

    \definecolor{mblue}{rgb}{0,0.447,0.741}
    \definecolor{mred}{rgb}{0.85,0.325,0.098}
    \definecolor{myellow}{rgb}{0.9290,0.6940,0.1250}
    \definecolor{mmagenta}{rgb}{1,0,1}
    \definecolor{mgreen}{rgb}{0.4460,0.6740,0.1880}
    \definecolor{mgrey}{rgb}{0.6,0.6,0.6}
    \definecolor{mpurple}{rgb}{0.4940, 0.1840, 0.5560}

    \definecolor{matbluel}{rgb}{0,0.6732,1}
    \definecolor{matyeld}{rgb}{0.2787,0.2082,0.0375}

    \tikzset{cross/.style={cross out, draw=black, minimum size=2*(#1-\pgflinewidth), inner sep=0pt, outer sep=0pt}, cross/.default={1pt}}
    
    \usetikzlibrary{shapes}

\newcommand{\blueline}{\raisebox{2pt}{\tikz{\draw[-,matblue1,solid,line width = 0.9pt](0,0) -- (3mm,0);}}}

\newcommand{\yelline}{\raisebox{2pt}{\tikz{\draw[-,matyel1,solid,line width = 0.9pt](0,0) -- (3mm,0);}}}
\newcommand{\purline}{\raisebox{2pt}{\tikz{\draw[-,matpur1,solid,line width = 0.9pt](0,0) -- (3mm,0);}}}
\newcommand{\greenline}{\raisebox{2pt}{\tikz{\draw[-,matgre1,solid,line width = 0.9pt](0,0) -- (3mm,0);}}}

\newcommand{\thicklblueline}{\raisebox{0pt}{\tikz{\draw[-,matbluel,solid,line width = 4pt](0,0) -- (3mm,0);}}}
\newcommand{\thickdyeline}{\raisebox{0pt}{\tikz{\draw[-,matyeld,solid,line width = 3pt](0,0) -- (3mm,0);}}}
\newcommand{\thickblueline}{\raisebox{1.3pt}{\tikz{\draw[-,matblue1,solid,line width = 1](0,0) -- (3mm,0);}}}
\newcommand{\thickyelline}{\raisebox{1.3pt}{\tikz{\draw[-,matyel1,solid,line width = 1](0,0) -- (3mm,0);}}}

\usepackage{tikz}
\usepackage{xcolor}

\newlength{\boxplotlinewidth} 
\newcommand{\boxploticonyel}{
    \begin{tikzpicture}[baseline=-0.2ex, scale=0.15, inner sep=0pt, outer sep=0pt]
        \path[use as bounding box] (-0.01,-0.6) rectangle (0.01,1.3); 
        \definecolor{iconcolor}{named}{matyel1}
        \draw[iconcolor, line width=\boxplotlinewidth] (-0.2,1.2) -- (0.2,1.2);
        \draw[iconcolor, line width=\boxplotlinewidth] (-0.2,-0.5) -- (0.2,-0.5);
        \draw[iconcolor, line width=\boxplotlinewidth] (0,0) -- (0,-0.5);
        \draw[iconcolor, line width=\boxplotlinewidth] (0,0.7) -- (0,1.2);
        \draw[iconcolor, line width=\boxplotlinewidth] (-0.3,0) -- (0.3,0);
        \draw[iconcolor, line width=\boxplotlinewidth] (-0.3,0.7) -- (0.3,0.7);
        \draw[iconcolor, line width=\boxplotlinewidth] (-0.3,0) -- (-0.3,0.7);
        \draw[iconcolor, line width=\boxplotlinewidth] (0.3,0) -- (0.3,0.7);
        \draw[iconcolor, line width=\boxplotlinewidth] (-0.3,0.35) -- (0.3,0.35);
    \end{tikzpicture}
}

\newcommand{\boxploticonblue}{
    \begin{tikzpicture}[baseline=-0.2ex, scale=0.15, inner sep=0pt, outer sep=0pt]
        \path[use as bounding box] (-0.01,-0.6) rectangle (0.01,1.3); 
        \definecolor{iconcolor}{named}{matblue1}
        \draw[iconcolor, line width=\boxplotlinewidth] (-0.2,1.2) -- (0.2,1.2);
        \draw[iconcolor, line width=\boxplotlinewidth] (-0.2,-0.5) -- (0.2,-0.5);
        \draw[iconcolor, line width=\boxplotlinewidth] (0,0) -- (0,-0.5);
        \draw[iconcolor, line width=\boxplotlinewidth] (0,0.7) -- (0,1.2);
        \draw[iconcolor, line width=\boxplotlinewidth] (-0.3,0) -- (0.3,0);
        \draw[iconcolor, line width=\boxplotlinewidth] (-0.3,0.7) -- (0.3,0.7);
        \draw[iconcolor, line width=\boxplotlinewidth] (-0.3,0) -- (-0.3,0.7);
        \draw[iconcolor, line width=\boxplotlinewidth] (0.3,0) -- (0.3,0.7);
        \draw[iconcolor, line width=\boxplotlinewidth] (-0.3,0.35) -- (0.3,0.35);
    \end{tikzpicture}
}

\title{\LARGE \bf
Robust Commutation Design: Applied to Switched Reluctance Motors
}

\author{Max van Meer$^{1}$, Gert Witvoet$^{1,2}$, Tom Oomen$^{1,3}$
\thanks{$^{1}$Max van Meer (e-mail: m.v.meer@tue.nl), Gert Witvoet, and Tom Oomen are with the Control Systems Technology section, Department of Mechanical Engineering, Eindhoven University of Technology, The Netherlands. This work is part of the research programme VIDI with project number 15698, which is (partly) financed by the Netherlands Organisation for Scientific Research (NWO). In addition, this research has received funding from the ECSEL Joint Undertaking under grant agreement 101007311 (IMOCO4.E). The Joint Undertaking receives support from the European Union's Horizon 2020 research and innovation programme.}
\thanks{$^{2}$Gert Witvoet is also with the Department of Optomechatronics, TNO, Delft, The Netherlands.}
\thanks{$^{3}$Tom Oomen is also with the Delft Center for Systems and Control, Delft University of Technology, Delft, The Netherlands.}
}

\let\labelindent\relax
    \usepackage{enumitem}
    
    \renewcommand{\baselinestretch}{0.99}
    \usepackage{eso-pic}
  \AddToShipoutPictureBG*{%
  \AtPageUpperLeft{%
  \setlength\unitlength{0.5in}%
  \hspace*{\dimexpr0.5\paperwidth\relax}
  \makebox(0,-1.75)[c]{
  \begin{tabular}{c c}
  Max van Meer, 
  Robust Commutation Design: Applied to Switched Reluctance Motors, \\
  Accepted for 
  {\em 22nd European Control Conference},
  Stockholm, Sweden, 2024,
  uploaded to arXiv \today \\
  \end{tabular}}}}

\begin{document}

\maketitle
\thispagestyle{empty}
\pagestyle{empty}

\begin{abstract}
Switched Reluctance Motors (SRMs) are cost-effective electric actuators that utilize magnetic reluctance to generate torque, with torque ripple arising from unaccounted manufacturing defects in the rotor tooth geometry.
This paper aims to design a versatile, resource-efficient commutation function for accurate control of a range of SRMs, mitigating torque ripple despite manufacturing variations across SRMs and individual rotor teeth.
The developed commutation function optimally distributes current between coils by leveraging the variance in the torque-current-angle model and is designed with few parameters for easy integration on affordable hardware. 
Monte Carlo simulations and experimental results show a tracking error reduction of up to 31\% and 11\%, respectively. 
The developed approach is beneficial for applications using a single driver for multiple systems and those constrained by memory or modeling effort, providing an economical solution for improved tracking performance and reduced acoustic noise.
\end{abstract}
\section{Introduction}
Switched Reluctance Motors (SRMs) have gained industrial interest due to their compelling advantages in energy efficiency, simplicity of design, and lack of permanent magnets, especially for low-cost applications that involve mass production~\cite{Miller1993,Kramer2020}. The working principle of SRMs involves the controlled switching of currents to different coils to produce magnetic attraction, a process that, if carried out imperfectly, can give rise to torque ripple.

Torque ripple is a common challenge in the implementation of SRMs, and it has a range of different causes such as sampling~\cite{VANMEER2022302}, magnetic hysteresis~\cite{Katalenic2013} and magnetic saturation~\cite{Boumaalif2023}, and the most important is imperfect commutation. The mechanism through which currents are applied to different coils to produce torque is specified by a user-defined commutation function \cite{VANMEER2022302,Xue2009a,Wang2016,Vujicic2012}. The design of such a function relies on a model of the torque-current-angle (TCA) relationship of the SRM. Any mismatch between this simplified model and the true system leads to a position-dependent error between the desired torque and the achieved torque, degrading the tracking performance of the system and introducing acoustic noise. Reasons for model mismatch include manufacturing tolerances and assembly variations. 

\begin{figure}\vspace{5pt}
    \centering
    \includegraphics[width=0.7\linewidth]{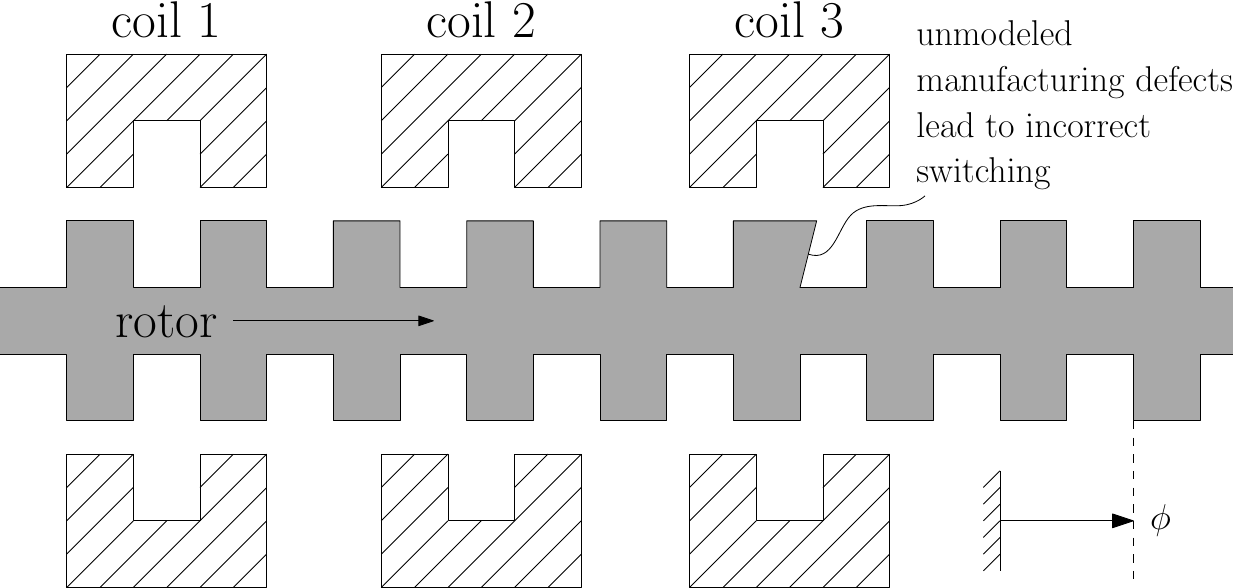}
    \caption{Schematic overview of an SRM with three coils. Sequentially applying currents to the coils attracts rotor teeth, generating torque. When control designs involve commutation functions that rely on incorrect or incomplete models, torque ripple occurs, degrading tracking performance.}\label{fig:srm}
    \end{figure}

The rotor of an SRM features a large number of teeth, each of which is slightly different due to imperfections in the manufacturing process, resulting in significant tooth-dependency in the TCA relationship of an SRM~\cite{Mooren2024}, see Figure~\ref{fig:srm}. Similarly, in mass production, manufacturing tolerances lead to variations in the rotor teeth across different SRMs as well. Modeling all these variations requires a tremendous effort when compared to using a commutation function which is designed using a low-order model of only one `average' tooth. Indeed, in low-cost applications that involve mass production, it is economically desirable to ship each SRM with the same driver, each of which has sufficient memory to store the TCA relationship of only one average tooth. 

\begin{figure}
        \centering
        \includegraphics[width=0.7\linewidth]{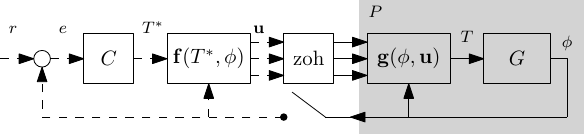}
        \caption{Control scheme for an SRM $P$: The SRM's nonlinear dynamics are linearized using a commutation function $\mathbf{f}$ to achieve $\hat{\mathbf{g}}\mathbf{f}=\pm 1$, enabling the use of a linear feedback controller $C(z)$. Solid lines and dashed lines depict continuous-time and discrete-time signals, respectively.}\label{fig:controlscheme}
        \end{figure}

Torque ripple is inevitable when variations across teeth or SRMs are ignored for economic reasons, and yet the magnitude of the torque ripple is affected by the choice of the commutation function, see Figure~\ref{fig:controlscheme}. At the same time, there is a high degree of freedom in the design of commutation functions~\cite{VANMEER2022302}, because multiple coil currents together lead to a single torque on the rotor. It is therefore hypothesized that there exists a low-order commutation function that mitigates torque ripple across many teeth and SRMs while relying on a low-order model of only one average tooth.  

Measurement data should be exploited to understand how manufacturing defects affect the TCA relationship of SRMs. In,~\cite{Meer2023b}, a data-driven identification approach is presented that yields accurate TCA models for use in commutation design, without relying on separate dedicated torque sensors. Importantly, this method not only yields an estimated model but also the variance of the model parameters. This variance reflects the tooth-by-tooth variations, or, when data from different SRMs is used, SRM-to-SRM variations. 

Although existing approaches to commutation function design are effective in the presence of a perfect model of every single tooth for each SRM, these methods lead to significant torque ripple in case of model mismatch. To require detailed, high-order models of SRMs defeats the purpose of using an SRM in many low-cost applications, not only because this requires expensive modeling effort, but also because it leads to high requirements on the driver hardware. Therefore, this paper aims to develop a universal, resource-efficient commutation function that leverages the model variance to reduce torque ripple across different rotor teeth and SRMs. This is achieved by using the parameter variance matrix resulting from~\cite{Meer2023b} to pose an optimization problem that minimizes the expected norm of the anticipated torque ripple. 
The contributions of this paper are threefold:\begin{enumerate}[label={C\arabic*:}]
    \item A method for the design of robust commutation functions is developed. Through convex optimization, torque ripple is minimized despite manufacturing defects. 
    \item Monte Carlo simulations show that a single robust commutation function leads to increased tracking performance accross various mass-produced SRMs. 
    \item Experimental results show that robust commutation functions effectively reduce torque ripple due to variations in individual SRM teeth. 
\end{enumerate}
This paper is structured as follows. In Section~\ref{sec:problem}, the problem formulation is given. Next, in Section~\ref{sec:robust}, the design method is detailed. Subsequently, simulation results and experimental results are presented in Section~\ref{sec:simulation} and~\ref{sec:experiment} respectively. Finally, conclusions are drawn in Section~\ref{sec:conclusions}.

\section{Problem formulation}\label{sec:problem}
This section presents the problem formulation, covering the nonlinear dynamics and control of SRMs, followed by defining the commutation design problem.
\subsection{Dynamics of Switched Reluctance Motors}
SRMs feature a nonlinear relationship between torque, current, and rotor angle. Neglecting magnetic saturation, an SRM with $n_t$ teeth and $n_c$ coils is modeled as:
\begin{equation}
T_c(\phi,i_c) = \frac{1}{2} \frac{\textnormal{d} L_c(\phi)}{\textnormal{d} \phi} i_c^2,
\end{equation}
where $T_c$ represents the torque applied to the rotor by magnetizing coil $c\in \{1,\ldots,n_c\}$ with a current $i_c$. $L_c(\phi)$ denotes the phase inductance, which varies periodically with the rotor position $\phi$, having a spatial period of $\frac{2\pi}{n_t}$. Both the torque $T$ and inductances $L_c$ are unmeasured and hence unknown. The total torque applied to the rotor at time $t$ is
\begin{equation}\label{eq:nonlin_plant}
T(t) = \mathbf{g}(\phi(t)) \mathbf{u}(t),
\end{equation}
where $\mathbf{g}(\phi)$ is defined as:
\begin{align}
\mathbf{g}(\phi) & :=\frac{1}{2}\frac{\textnormal{d}}{\textnormal{d}\phi}[L_1(\phi(t)),\ldots,L_{n_c}(\phi(t))],
\end{align}
and $\mathbf{u}(t)$ represents the squared coil currents:
\begin{equation}
\mathbf{u}{(t)} := [i_1^2{(t)},\ldots,i_{n_c}^2{(t)}]^\top.
\end{equation}
The following section addresses how a desired torque $T^*$ is realized, using a model $\hat{\mathbf{g}}\approx \mathbf{g}$.

\subsection{Commutation: linearization of SRM dynamics} 
To achieve a desired torque $T^*$ in SRMs, we invert the nonlinear TCA relationship~\eqref{eq:nonlin_plant} through a commutation function $\mathbf{u}=\mathbf{f}(\phi, T^*)$ as follows. First, $\mathbf{f}$ is structured as \begin{equation}\label{eq:f}
    \mathbf{f}(\phi,T^*) := \begin{cases}
    \mathbf{f}^+(\phi) T^* & T^*\geq 0,\\
    -\mathbf{f}^-(\phi) T^* & T^*<0,
    \end{cases}
    \end{equation}
where $\mathbf{f}^+(\phi),\mathbf{f}^-(\phi): \mathbb{R}\to \mathbb{R}^{n_c}$ satisfy \begin{equation}\label{eq:requirements}
    \begin{aligned}
    \hat{\mathbf{g}}(\phi)\mathbf{f}^+(\phi) &\approx 1, &\quad \hat{\mathbf{g}}(\phi)\mathbf{f}^-(\phi) &\approx -1,
    \end{aligned}
    \end{equation}
    such that when $\mathbf{f}^+$ or $\mathbf{f}^-$ is multiplied with a desired torque $T^*$, the resulting currents lead to $T\approx T^*$, see~\eqref{eq:nonlin_plant}.
Moreover, as $\mathbf{f}$ produces squared currents, it must be ensured that \begin{equation}
    \begin{aligned}
    \mathbf{f}^+(\phi) &\geq \mathbf{0}, &\quad \mathbf{f}^-(\phi)&\geq \mathbf{0}.
    \end{aligned}
    \end{equation}
    It is important to note that multiple functions $\mathbf{f}(\phi, T^*)$ satisfy these requirements due to $\mathbf{g}$ and $\mathbf{f}$ being row and column vector functions, respectively. By applying the control law
    \begin{equation}\label{eq:controllaw}
    \mathbf{u}(t) = \mathbf{f}(\phi(t),T^*(t)),
    \end{equation}
    the squared currents for each coil to achieve the desired torque $T^*$ are determined, as visualized in Figure~\ref{fig:controlscheme}. Combined with~\eqref{eq:nonlin_plant}, the resulting torque is:
    \begin{equation}\label{eq:truetorque}
    T(\phi(t)) = \mathbf{g}(\phi(t)) \mathbf{f}(\phi(t),T^*(t)),
    \end{equation}
    or equivalently,\begin{equation}\label{eq:truetorque_b}
        \begin{aligned}
            T(\phi(t)) =&b^{\pm}\left(\phi(t)\right)T^* \operatorname{sign}(T^*),\\
            b^{\pm}(\phi) :=& \mathbf{g}(\phi) \mathbf{f}^{\pm}(\phi),\\
            \hat{b}^{\pm}(\phi) :=& \hat{\mathbf{g}}(\phi)\mathbf{f}^{\pm}(\phi).
            \end{aligned}
    \end{equation}
    Here, $\pm:=\text{sign}(T^*)$, e.g., $\mathbf{f}^{\pm}$ refers to either $\mathbf{f}^{+}$ or $\mathbf{f}^{-}$, depending on the sign of $T^*$. Additionally, $b^{\pm}(\phi)$ denotes the true relative torque mismatch $T/T^*$ while $\hat{b}^{\pm}(\phi)$ represents the expected mismatch, which is $\pm 1$ as per design, see~\eqref{eq:requirements}.

    Ideally, if $\hat{\mathbf{g}}(\phi)=\mathbf{g}(\phi)$, and $\mathbf{f}$ meets the requirements in~\eqref{eq:requirements}, then $T=T^*$. However, if $\hat{\mathbf{g}}(\phi)\neq \mathbf{g}(\phi)$, torque ripple occurs, degrading the tracking performance~\cite{Gan2018}. The next subsection defines the problem of designing commutation functions $\mathbf{f}$ that counteract such model imperfections.
\subsection{Problem definition} 
Our purpose is to design a commutation function $\mathbf{f}(\phi,T^*)$, structured as in~\eqref{eq:f}, that mitigates torque ripple arising from modeling errors $\hat{\mathbf{g}}(\phi)\neq\mathbf{g}(\phi)$. Specifically, the objective is to minimize the true relative torque error \begin{equation}\begin{aligned}\label{eq:torque_error}
    \varepsilon_T^{\pm}(\phi) &:= {b}^{\pm}(\phi) \mp 1,\\
    &= \mathbf{g}(\phi) \mathbf{f}^{\pm}(\phi) \mp 1,
\end{aligned}
\end{equation}
where $\mp:=-\text{sign}(T^*)$. We consider the following cost:
\begin{equation}\begin{aligned}
    \mathcal{J}&= \|\varepsilon_T^{+}(\phi)\|_{2,\Phi}^2 + \|\varepsilon_T^{-}(\phi)\|_{2,\Phi}^2,\\
    \text{with }\Phi&=\{\phi\mid 0\leq \phi < {2\pi}\}.
\end{aligned}
    \end{equation}
Since $\mathbf{g}$ is unknown, the torque error $\varepsilon_T^{\pm}(\phi)$ is unknown, so this cost function cannot be evaluated directly. 
Therefore, it is assumed that the model $\hat{\mathbf{g}}(\phi,\boldsymbol{\theta})$ with parameters $\boldsymbol{\theta}\in \mathbb{R}^{n_{\boldsymbol{\theta}}}$ is probabilistic, indicating that some information about the modeling errors is available. More precisely, $\boldsymbol{\theta}$ is a multivariate Gaussian with \begin{equation}\label{eq:random_theta}
    \boldsymbol{\theta}\sim\mathcal{N}(\hat{\boldsymbol{\theta}},\boldsymbol{\Sigma}_{\boldsymbol{\theta}}).
\end{equation}
Identification of $\hat{\boldsymbol{\theta}}$ and $\boldsymbol{\Sigma}_{\boldsymbol{\theta}}$ is addressed in~\cite{Meer2023b}. In this paper, $\hat{\boldsymbol{\theta}}$ and $\boldsymbol{\Sigma}_{\boldsymbol{\theta}}$ are assumed to be known, with $\hat{\boldsymbol{\theta}}$ generally unequal to $\boldsymbol{\theta}_{\text{true}}$ and $\boldsymbol{\Sigma}_{\boldsymbol{\theta}}$ positive definite. The objective is then to minimize the following cost function: \\
\begin{equation}\begin{aligned}
    \hat{\mathcal{J}}&= \mathbb{E}\left[\|\hat{\varepsilon}_T^{+}(\phi)\|_{2,\Phi'}^2 + \|\hat{\varepsilon}_T^{-}(\phi)\|_{2,\Phi'}^2\right],\\
    \text{with }\Phi'&=\{\phi\mid 0\leq \phi < {2\pi}/n_t\},
\end{aligned}
    \end{equation}
    where $\Phi'$ spans only a single tooth because we desire a resource-efficient commutation function that is the same for every tooth, using a model $\hat{\mathbf{g}}(\phi,\boldsymbol{\theta})$ that is also tooth-invariant, albeit probabilistic. 
In the next section, the model variance $\boldsymbol{\Sigma}_{\boldsymbol{\theta}}$ is exploited in the design of $\mathbf{f}$, to minimize torque ripple arising from modeling errors $\hat{\mathbf{g}}(\phi,\boldsymbol{\theta})\neq\mathbf{g}(\phi)$. 
\section{Robust commutation function design}\label{sec:robust}
This section covers the design of robust commutation functions, including the SRM model structure, torque ripple caused by model mismatch, and the optimization problem.
\subsection{Parametrization of commutation functions and dynamics}
The given SRM model, denoted by $\hat{\mathbf{g}}(\phi,\boldsymbol{\theta})$, as well as $\mathbf{f}^{\pm}(\phi,\boldsymbol{\alpha})$, are both  parametrized linearly in their parameters. The structure of the provided SRM model is given by
\begin{equation}\label{eq:model_g}
\hat{\mathbf{g}}^{\top}(\phi,\boldsymbol{\theta})=\boldsymbol{\psi}_g(\phi)\boldsymbol{\theta},
\end{equation}
where $\boldsymbol{\psi}_g:\mathbb{R}\to\mathbb{R}^{n_c \times n_{\theta}}$ serves as the basis of $\hat{\mathbf{g}}(\phi,\boldsymbol{\theta})$. 
Similarly, the commutation functions $\mathbf{f}^+$ and $\mathbf{f}^-$ are parametrized as follows. Both $\mathbf{f}^+$ and $\mathbf{f}^-$ are designed with $n_{\alpha}$ parameters per coil, i.e., $\alpha^{+}_{c,i}$ and $\alpha^{-}_{c,i}$ denote the $i^{\text{th}}$ parameter of coil $c\in\{1,\ldots n_c\}$ of commutation functions $\mathbf{f}^+$ and $\mathbf{f}^-$ respectively. We then define $\mathbf{f}^+$ and $\mathbf{f}^-$ as
\begin{equation}\label{eq:model_f}
\mathbf{f}^{\pm}(\phi,\boldsymbol{\alpha})=\boldsymbol{\psi}_f(\phi) \boldsymbol{\alpha}^{\pm}, 
\end{equation}
where $\boldsymbol{\alpha}^{\pm}=[{{\boldsymbol{\alpha}^{\pm}}^\top_{1}},\ldots,{{\boldsymbol{\alpha}^{\pm}}^\top_{c}},\ldots,{{\boldsymbol{\alpha}^{\pm}}^\top_{n_c}}]^\top$ stacks the parameters of all coils and $\boldsymbol{\alpha}_{c}^{\pm}=[{\alpha^{\pm}}^\top_{c,1},\ldots,{\alpha^{\pm}}^\top_{c,n_{\alpha}}]^\top$ are the parameters of coil $c$. Moreover, $\boldsymbol{\alpha}=[\boldsymbol{\alpha}^{+\top}, \boldsymbol{\alpha}^{-\top}]^\top$, and
\begin{equation}
    \boldsymbol{\psi}_f(\phi) = \mathbf{I}_{n_c} \otimes \boldsymbol{\gamma}(\phi),
\end{equation}
with $\boldsymbol{\gamma}(\phi)$ the basis for $\mathbf{f}^{\pm}$ and $\otimes$ the Kronecker product. In the current paper, commutation functions are defined by the following basis. Given a grid $\boldsymbol{\phi}_{\gamma}\in\mathbb{R}^{n_{\alpha}}$ with rotor positions $\phi_{\gamma,i}$ spaced between 0 and $2\pi/n_t$, we define 
\begin{equation}\label{eq:gamma}
    \boldsymbol{\gamma}(\phi)=\begin{bmatrix}
        k(\rho_1(\phi)),\ldots,k(\rho_{n_{\alpha}}(\phi))
    \end{bmatrix},
\end{equation}
where \begin{equation}
    \rho_i(\phi) = \frac{1}{\ell}\sqrt{(\mathbf{x}_{1,i}-\mathbf{x}_{2}(\phi))^{\top}(\mathbf{x}_{1,i}-\mathbf{x}_{2}(\phi))}
\end{equation}
with length scale $\ell>0$, and \begin{equation}\begin{aligned}\label{eq:sincos}
    \mathbf{x}_{1,i}&=\begin{bmatrix}\sin(\phi_{\gamma,i} n_t) \\ \cos(\phi_{\gamma,i}n_t)\end{bmatrix},&\mathbf{x}_{2}(\phi)&=\begin{bmatrix}\sin(\phi n_t) \\ \cos(\phi n_t)\end{bmatrix}.
\end{aligned}
\end{equation}
Moreover, the kernel function $k$ in~\eqref{eq:gamma} is given by \begin{equation}\begin{aligned}\label{eq:kernel}
    k(\rho)&=\exp \left(-{\sqrt{2 \mu+1} \rho}\right) \frac{\mu !}{(2 \mu) !} \\ &\cdot\sum_{n=0}^{\mu} \frac{(\mu+n) !}{n !(\mu-n) !}\left({2 \sqrt{2 \mu+1} \rho}\right)^{\mu-n}.
    \end{aligned}
    \end{equation}
    This model structure draws inspiration from Gaussian Process (GP) regression, where $k$ is recognized as a Matèrn kernel. These choices are new in the design of commutation functions. The kernel enforces periodicity and smoothness and is very successful in GP regression \cite{VANMEER2022302}. Specifically,~\eqref{eq:sincos} ensures that $\mathbf{f}^{\pm}$ exhibits periodic behavior corresponding to the spatial period of a rotor tooth. Moreover, the designer can control the smoothness of $\mathbf{f}^{\pm}$ by adjusting the length scale $\ell$ or the parameter $\mu\in\mathbb{N}$.
\subsection{Torque ripple quantification in commutation design}
The objective is designing $\mathbf{f}^{\pm}(\phi,\boldsymbol{\alpha})$ to minimize the expected torque ripple over all possible realizations of the random model $\hat{\mathbf{g}}$. To achieve this, we first define the estimated relative torque error $\hat{\varepsilon}_T^{\pm}(\phi,\boldsymbol{\theta},\boldsymbol{\alpha})\approx {\varepsilon}_T^{\pm}(\phi,\boldsymbol{\alpha})$ as \begin{equation}\begin{aligned}\label{eq:torque_error_est}
    \hat{\varepsilon}_T^{\pm}(\phi,\boldsymbol{\theta},\boldsymbol{\alpha}) &:= \hat{b}^{\pm}(\phi) \mp 1,\\
    &= \hat{\mathbf{g}}(\phi,\boldsymbol{\theta}) \mathbf{f}^{\pm}(\phi,\boldsymbol{\alpha}) \mp 1.
\end{aligned}
\end{equation}
Substitution of~\eqref{eq:model_g} and~\eqref{eq:model_f} yields \begin{equation}\label{eq:hatvareps}
    \hat{\varepsilon}_T^{\pm}(\phi,\boldsymbol{\theta},\boldsymbol{\alpha}) = \boldsymbol{\theta}^{\top}\boldsymbol{\psi}_g^{\top}(\phi)\boldsymbol{\psi}_f(\phi)\boldsymbol{\alpha}^{\pm}\mp 1.
\end{equation}
For reasons that become apparent later, a vector $\hat{\boldsymbol{\varepsilon}}_T$ is defined that stacks $\hat{\varepsilon}_T^{\pm}(\phi)$ as follows: 
     \begin{equation}
        \hat{\boldsymbol{\varepsilon}}_T=[
            \hat{{\varepsilon}}_T^{+}(\phi_{\varepsilon,1}),\ldots,\hat{\varepsilon}_T^{+}(\phi_{\varepsilon,N}), \hat{{\varepsilon}}_T^{-}(\phi_{\varepsilon,1}),\ldots,\hat{\varepsilon}_T^{-}(\phi_{\varepsilon,N})
        ]^{\top}.
    \end{equation}
Here, $\boldsymbol{\phi}_{\varepsilon}\in \mathbb{R}^N$ is an evenly spaced grid of rotor angles $\phi_{\varepsilon,i}$ between 0 and $2\pi/n_t$. The expression for the vector $\hat{\boldsymbol{\varepsilon}}_T$ of estimated torque errors is then given by \begin{equation}\label{eq:torquevec}
    \hat{\boldsymbol{\varepsilon}}_T = \mathbf{X}(\boldsymbol{\alpha}) \boldsymbol{\theta} + \begin{bmatrix}
        -\mathbf{1}_N\\\mathbf{1}_N
    \end{bmatrix},
\end{equation}
with \begin{equation}
    \mathbf{X}(\boldsymbol{\alpha})=\mathbf{F}(\boldsymbol{\alpha}) \boldsymbol{\Psi}_g,
\end{equation}
where $\boldsymbol{\Psi}_g=\mathbf{1}_2\otimes[\boldsymbol{\psi}_g^\top(\phi_{\varepsilon,1}),\ldots,\boldsymbol{\psi}_g^\top(\phi_{\varepsilon,N})]^\top$ and \begin{equation}\begin{aligned}
    \mathbf{F}(\boldsymbol{\alpha}) 
    &=\sum_{i=1}^{2N} \mathbf{E}_{ii}\otimes (\boldsymbol{\alpha}^{\Omega_i\top}\boldsymbol{\psi}_f^\top(\phi_{\varepsilon,i})),\\
    \text{with }\Omega_i &:= \begin{cases}
        + & 1 \leq i \leq N,\\
        - & N < i \leq 2N.
        \end{cases}
\end{aligned}
\end{equation}
Here, $\mathbf{E}_{ii}\in\mathbb{B}^{2N\times 2N}$ denotes a matrix unit, which has only one nonzero entry at the $i^{\text{th}}$ row and column.\\
In summary,~\eqref{eq:torquevec} yields a vector $\hat{\boldsymbol{\varepsilon}}_T$ of estimated torque errors, evaluated on a grid $\boldsymbol{\phi}_{\varepsilon}$. Since this vector is linear in the random SRM model parameters $\boldsymbol{\theta}$, see~\eqref{eq:random_theta}, we can write \begin{equation}\begin{aligned}\label{eq:distribution_e}
    \hat{\boldsymbol{\varepsilon}}_T &\sim \mathcal{N}\left(\boldsymbol{\mu}_{\varepsilon},\boldsymbol{\Sigma}_{\varepsilon}\right),\\
    \boldsymbol{\mu}_{\varepsilon}&=       \mathbf{X}(\boldsymbol{\alpha})\boldsymbol{\hat{\theta}}+\begin{bmatrix}
        -\mathbf{1}_N\\\mathbf{1}_N
    \end{bmatrix},\\
    \boldsymbol{\Sigma}_{\varepsilon} &=\mathbf{X}(\boldsymbol{\alpha})\boldsymbol{\Sigma}_{\boldsymbol{\theta}} \mathbf{X}^\top(\boldsymbol{\alpha}).
\end{aligned}
\end{equation}
In this way, the estimated torque error is expressed in terms of the variance $\boldsymbol{\Sigma}_{\boldsymbol{\theta}}$ of the SRM model parameters $\boldsymbol{\theta}$. Two key observations can be made from~\eqref{eq:distribution_e}. First, more uncertainty in the model parameters leads to a larger variance of $\hat{\boldsymbol{\varepsilon}}_T$, i.e., potentially more torque ripple, even if $\boldsymbol{\alpha}$ is designed such that $\hat{\boldsymbol{\varepsilon}}_T$ is zero-mean. Second, the variance of $\hat{\boldsymbol{\varepsilon}}_T$ is quadratically dependent on $\boldsymbol{\alpha}$, indicating that $\boldsymbol{\alpha}$ can be tuned to obtain robust commutation functions $\mathbf{f}(\phi,\boldsymbol{\alpha})$ that have minimal variance $\boldsymbol{\Sigma}_{\varepsilon}$ of the expected torque ripple $\hat{\boldsymbol{\varepsilon}}_T$. The next section describes how this can be achieved. 
\subsection{Optimization problem formulation}
The objective is to find the optimal $\boldsymbol{\alpha}$ that minimizes the estimated torque ripple, $\hat{\boldsymbol{\varepsilon}}_T$. The cost function is defined as:
\begin{equation}
        \begin{aligned}
            \tilde{\mathcal{J}}(\boldsymbol{\alpha})=&\mathbb{E}\left[\|\hat{\varepsilon}_T^{+}(\phi,\boldsymbol{\theta},\boldsymbol{\alpha}^+)\|_{2,\Phi'}^2 + \|\hat{\varepsilon}_T^{-}(\phi,\boldsymbol{\theta},\boldsymbol{\alpha}^-)\|_{2,\Phi'}^2\right]\\
            =& \mathbb{E}\left[\int_{0}^{2\pi/n_t}(\hat{\varepsilon}_T^{+}(\phi,\boldsymbol{\theta},\boldsymbol{\alpha}^+))^2 +(\hat{\varepsilon}_T^{-}(\phi,\boldsymbol{\theta},\boldsymbol{\alpha}^-))^2 d\phi\right] \\
                   \approx & \frac{2\pi}{n_t N} \mathbb{E}\left[\|\hat{\boldsymbol{\varepsilon}}_T\|_2^2\right].
        \end{aligned}
        \end{equation}
        We henceforth drop the scaling factor $\frac{2\pi}{n_t N}$ as it does not affect the optimizer. Since $\hat{\boldsymbol{\varepsilon}}_T$ is random, see~\eqref{eq:distribution_e}, we have
\begin{equation}\label{eq:cost}\begin{aligned}
\hat{\mathcal{J}}(\boldsymbol{\alpha})=& \mathbb{E}\left[\hat{\boldsymbol{\varepsilon}}_T^\top \hat{\boldsymbol{\varepsilon}}_T\right]= \text{tr}\left(\mathbb{E}\left[\hat{\boldsymbol{\varepsilon}}_T^\top \hat{\boldsymbol{\varepsilon}}_T\right]\right)=\mathbb{E}\left[\text{tr}\left(\hat{\boldsymbol{\varepsilon}}_T \hat{\boldsymbol{\varepsilon}}_T^\top\right)\right]\\
=& \text{tr}(\boldsymbol{\Sigma}_{\varepsilon}) + \boldsymbol{\mu}_{\varepsilon}^\top \boldsymbol{\mu}_{\varepsilon},\\
=&\text{tr}\left(\mathbf{X}(\boldsymbol{\alpha})\boldsymbol{\Sigma}_{\boldsymbol{\theta}} \mathbf{X}^\top(\boldsymbol{\alpha})\right)+\boldsymbol{\hat{\theta}}^\top \mathbf{X}^\top(\boldsymbol{\alpha}) \mathbf{X}(\boldsymbol{\alpha})\boldsymbol{\hat{\theta}}\\
&-2\begin{bmatrix} 
    -\mathbf{1}_N^\top&\mathbf{1}_N^\top
\end{bmatrix}\mathbf{X}(\boldsymbol{\alpha})\boldsymbol{\hat{\theta}}+2N,
\end{aligned}
\end{equation}
see~\cite[Section 3.2b]{Mathai1992} for details. This results in a cost function that is quadratic in $\boldsymbol{\alpha}$, and convex since $\boldsymbol{\Sigma}_{\theta}$ is positive definite. Moreover, to ensure that the designed commutation functions yield positive squared currents for all gridded rotor angles, the following constraints are defined:
\begin{equation}\begin{aligned}
\left[\mathbf{f}^+(\phi_{\varepsilon,1},\boldsymbol{\alpha})^\top,\ldots\right.&,\mathbf{f}^+(\phi_{\varepsilon,N},\boldsymbol{\alpha})^\top,\\
&\left.\mathbf{f}^-(\phi_{\varepsilon,1},\boldsymbol{\alpha})^\top,\ldots,\mathbf{f}^-(\phi_{\varepsilon,N},\boldsymbol{\alpha})^\top\right]^\top \geq \mathbf{0},
\end{aligned}\end{equation}
which are expressed linearly in $\boldsymbol{\alpha}$ as
\begin{equation}
\mathbf{B}\boldsymbol{\alpha} \geq \mathbf{0},
\end{equation}
where \begin{equation}\begin{aligned}
        \mathbf{B} &= \sum_{i=1}^{2N} \left(\left(\mathbf{e}_i\otimes \mathbf{I}_{n_c}\right)\otimes \iota_i\right)\otimes \boldsymbol{\gamma}_f(\phi_{\varepsilon,i}),\\
        \text{with } \iota_i &:= \begin{cases}
            [1,\ 0] & 1 \leq i \leq N,\\
            [0,\ 1] & N < i \leq 2N.
            \end{cases}
\end{aligned}
\end{equation}
Here, $\mathbf{e}_i\in\mathbb{B}^{2N}$ is a unit vector with one nonzero entry at the $i^{\text{th}}$ element. Finally, the convex optimization problem for robust design of commutation functions is posed as \begin{equation}\label{eq:problem}
    \begin{array}{ll}
    \min_{\boldsymbol{\alpha}} & \tilde{\mathcal{J}}(\boldsymbol{\alpha}), \\
    \text {subject to } & \mathbf{B}\boldsymbol{\alpha} \geq \mathbf{0}.
    \end{array}
    \end{equation}
This problem with $2n_c n_{\alpha}$ design variables and $2 n_{c}N$ linear constraints is readily solved with a convex solver, e.g., CasADi, see~\cite{Andersson2019}. Note that in this framework, other terms can be easily included in the cost function and constraints, e.g., to mitigate power consumption, peak currents, maximum slew rates, or sampling-induced torque ripple~\cite{VANMEER2022302}. 
\section{Simulation results}\label{sec:simulation}
This section shows how the robust commutation design framework improves tracking despite model mismatches, using Monte Carlo simulations.
\subsection{Simulation setup}
\begin{figure}\vspace{5pt}
    \centering
    \includegraphics[width=\linewidth]{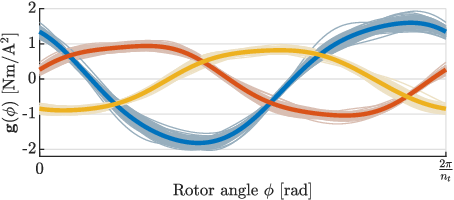}
    \caption{Torque-current-angle relationships $\mathbf{g}(\phi,\boldsymbol{\theta}_i)$ of the simulated SRMs ($\lambda=1$). The average $\mathbf{g}(\phi,\boldsymbol{\theta}^\circ)$ is shown in bold.}
    \label{fig:all_g}
    \end{figure}

A series of $M=100$ different SRMs is considered, each featuring $n_t=131$ identical teeth and $n_c=3$ coils. The linear dynamics of each SRM are represented as:
\begin{equation}
G(s):=\frac{\phi(s)}{T(s)} = \frac{1}{s(s+1)},
\end{equation}
and a PID controller $C(s)$ is designed to have a bandwidth of 20 Hz. The nonlinear TCA relationship of the SRMs is \begin{equation}
\mathbf{g}^\top(\phi,\boldsymbol{\theta}_i)=\boldsymbol{\psi}_g(\phi)\boldsymbol{\theta}_i,\quad i\in\{1,\ldots,M\},
\end{equation}
where the variation across the SRMs follows from \begin{equation}\label{eq:SRM_sims}
    \boldsymbol{\theta}_i\sim\mathcal{N}({\boldsymbol{\theta}^\circ},\lambda\boldsymbol{\Sigma}_{\boldsymbol{\theta}}), 
\end{equation} with $\lambda>0$, $\boldsymbol{\theta}\in\mathbb{R}^{90}$, and the structure $\boldsymbol{\psi}_g$ comprises of radial basis functions, see Figure~\ref{fig:all_g}.
The rotor angle $\phi$ is sampled at a rate of 5 kHz. At this rate, sampling-induced torque ripple is negligible at only 1\% of the total torque ripple~\cite{VANMEER2022302}, i.e., the tracking error is approximately zero when a commutation function $\mathbf{f}_i$ is designed to perfectly invert $\mathbf{g}_i$.

In this simulation study, the true systems $\mathbf{g}_i$ are unknown, and a single probabilistic model $\hat{\mathbf{g}}$ with structure~\eqref{eq:model_g} and $\boldsymbol{\theta}\sim\mathcal{N}({\boldsymbol{\theta}^\circ},\lambda\boldsymbol{\Sigma}_{\boldsymbol{\theta}})$ is used. This unbiased model, with the same structure as the true systems, is used to design a single robust $\mathbf{f}$. In this example, we choose $\hat{\boldsymbol{\theta}}=\boldsymbol{\theta}^\circ$ to focus specifically on the effect of minimizing the variance term $\operatorname{tr}(\boldsymbol{\Sigma}_\varepsilon)$ in~\eqref{eq:cost}, but $\hat{\boldsymbol{\theta}}$ may be unequal to $\boldsymbol{\theta}^\circ$ in general. Because of the variation across SRMs, it is inevitable that each SRM will suffer from torque ripple. 
\subsection{Benchmark: conventional commutation functions}
The performance of the robust commutation function $\mathbf{f}$ is compared to the performance obtained using a commonly used conventional function $\mathbf{f}_{\text{conv}}$. This function is given by 
\begin{equation}\label{eq:fconv}\begin{aligned}
    \mathbf{f}_{\text{conv}, c}\left(\phi,T^*\right)=&\mathbf{f}_{\mathrm{TSF},c}\left(\phi+\frac{2 \pi(c-1)}{n_c},T^*\right) \\
    &\cdot\operatorname{sat}\left(1 / \hat{{g}}_{c}\left(\phi,\boldsymbol{\theta}^\circ\right)\right)T^*,\end{aligned}
    \end{equation}
    where $\mathbf{f}_{\text{conv}, c}$ denotes the $c^{\mathrm{th}}$ element of $\mathbf{f}_{\text{conv}}$. Additionally, the saturation function $\operatorname{sat}(x)$ is defined as: \begin{equation}
    \textnormal{sat}(x) :=\begin{cases}x_{\min}  & x < x_{\min}, \\
    {x}  & {x_{\min}\leq x \leq x_{\max},} \\
    x_{\max} & x > x_{\max}.\end{cases}
    \end{equation}Moreover, $\mathbf{f}_{\mathrm{TSF}}(\phi,T^*): \mathbb{R}\times \mathbb{R}\to\mathbb{R}^{n_c}$ represents a torque sharing function that distributes a desired torque to different coils. As detailed in~\cite{Wang2016}, $\mathbf{f}_{\mathrm{TSF}}(\phi,T^*)$ is designed to satisfy  \begin{equation}
    \sum_{c=1}^{n_c} \mathbf{f}_{\mathrm{TSF},c}(\phi,T^*) = \begin{cases}
        1 & T^* \geq 0,\\
        -1 & T^* < 0.
    \end{cases}
    \end{equation} At values of $\phi$ where ${g}_c(\phi)=0$, $\mathbf{f}_{\mathrm{TSF},c}(\phi,T^*)=0$ by design, such that~\eqref{eq:fconv} is well defined for all $\phi$. 
\subsection{Results and analysis}
The developed framework is applied as follows. $\mathbf{f}$ is parametrized as in~\eqref{eq:f} and~\eqref{eq:model_f}, with $n_\alpha=50$, $\ell=0.3$ and $\mu=3$. With $N=100$, problem \eqref{eq:problem} has 300 design variables and 600 constraints. Using CasADi with MATLAB on a personal computer, the problem is solved in 300 s. 
\begin{figure}\vspace{5pt}
    \centering
    \includegraphics[width=\linewidth]{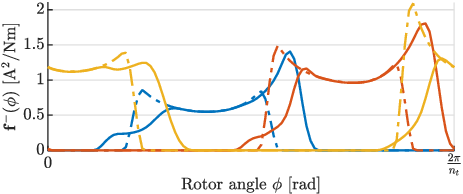}
    \caption{The developed robust commutation functions $\mathbf{f}^-$ (solid) and conventional functions $\mathbf{f}_{\text{conv}}^-$ (dot-dashed). The developed commutation functions exhibit much more overlap, leading to careful switching of the currents in the face of model uncertainty.}\label{fig:f}
    \end{figure}

The resulting function is shown in Figure~\ref{fig:f}. For brevity, we focus on $\mathbf{f}^-$ only. The resulting $\mathbf{f}^-$ exhibits more overlap than $\mathbf{f}_{\text{conv}}$, which can be interpreted as switching the currents carefully: near the angles $\phi$ where the currents are switched to the next coil, a sharp change in current at a slightly incorrect angle induces significant torque ripple. The conventional $\mathbf{f}_{\text{conv}}$ is designed purely for minimal power given some constraints on the slew rate and thus switches more abruptly, inducing torque ripple. 
\begin{figure}\vspace{5pt}
    \centering%
    \includegraphics[width=\linewidth]{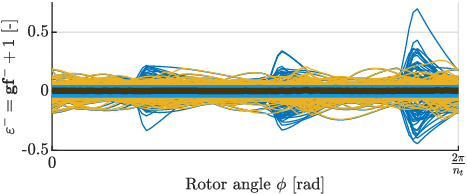}
    \caption{Torque ripple of all SRMs. Commutation functions $\mathbf{f}^-$ and $\mathbf{f}_{\text{conv}}$ are designed to invert $\mathbf{g}(\phi,\boldsymbol{\theta}^\circ)$ so the torque ripple is expected to be low with both $\mathbf{f}^-$ (\protect\thickdyeline) and $\mathbf{f}_{\text{conv}}$ (\protect\thicklblueline) for the average SRM. However, since each simulated SRM is different with $\boldsymbol{\theta}_i\neq \boldsymbol{\theta}^\circ$, each SRM exhibits torque ripple. The robust $\mathbf{f}^-$ exhibits significantly less torque ripple (\protect\thickyelline) than $\mathbf{f}_{\text{conv}}$ (\protect\thickblueline) because the model uncertainty is taken into account in the design.}\label{fig:b}
    \end{figure}

To see how $\mathbf{f}$ mitigates torque ripple, consider Figure~\ref{fig:b}. Both $\mathbf{f}$ and $\mathbf{f}_{\text{conv}}$ are designed to have approximately zero torque error for the nominal model $\hat{\mathbf{g}}(\phi,\boldsymbol{\theta}^\circ)$. However, when an SRM with dynamics $\mathbf{g}(\phi,\boldsymbol{\theta}_i)$ has parameters $\boldsymbol{\theta}_i$ ever so slightly different from $\boldsymbol{\theta}^\circ$, significant torque ripple occurs: up to 70\%. Indeed, $\mathbf{f}$ leads to significantly less torque ripple than $\mathbf{f}_{\text{conv}}$, indicating that the designed commutation functions truly are more robust to model uncertainty. Note that torque ripple is always expected when $\hat{\mathbf{g}}\neq {\mathbf{g}}$.

\begin{figure}\vspace{5pt}
    \centering
    \includegraphics[width=\linewidth]{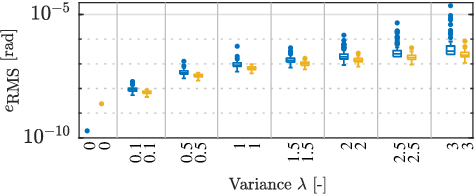}
    \caption{RMS tracking error as a function of the variance $\lambda$, where higher values of $\lambda$ corresponds to more model mismatch. The robust commutation functions $\mathbf{f}$ result in a smaller error (\protect\boxploticonyel) than the conventional functions $\mathbf{f}_{\text{conv}}$ (\protect\boxploticonblue) when even the slightest modeling errors are present.}\label{fig:e_var}
    \end{figure}

Next, four closed-loop simulations are carried out for all $M$ SRMs. First, a reference is tracked with a constant velocity of 0.3 teeth/s for a duration of 5 teeth, using $\mathbf{f}_{\text{conv}}$. Next, the simulation is repeated using $\mathbf{f}$, and finally, both simulations are repeated in the other direction. 
We define
\begin{equation}
    \begin{aligned}
    e_{\text{RMS}}^{\pm} &:= \sqrt{{\frac{1}{N_f-N_s}}\sum_{k=N_s}^{N_f} (e^{\pm}(t_k))^2},\\
    e_{\text{RMS}} &:= \sqrt{\frac{1}{2}\left((e_{\text{RMS}}^{+})^2+(e_{\text{RMS}}^{-})^2\right)},
    \end{aligned}
\end{equation}
with $N_s$ corresponding to the sample at which the rotor is at the second last tooth, and $N_f$ the last sample. Robust commutation reduces the median $e_{\text{RMS}}^{-}$ from 91 nrad to 63 nrad (-31\%) for $\lambda=1$. The maximum reduction is 84\% in case of particularly large model mismatch.
Finally, we study how performance depends on the variance magnitude $\lambda$, as per Equation~\eqref{eq:SRM_sims}. For $\lambda=0$, a single SRM with $\mathbf{g}=\mathbf{g}(\phi,\boldsymbol{\theta}^\circ)$ is simulated. For other $\lambda$ values, $M=100$ models $\boldsymbol{\theta}_i$ are sampled from $\boldsymbol{\theta}$ using the same seed, resulting in $\hat{\mathbf{g}}_i$ differing by a scaling factor across $\lambda$ values, shown in Figure~\ref{fig:e_var}. At $\lambda=0$, there is no model mismatch and conventional commutation functions yield lower error due to their parametrization, see~\eqref{eq:fconv}. Conversely, even with minimal variance ($\lambda=0.1$) robust commutation significantly improves performance, as illustrated in Figure~\ref{fig:e_var}. For substantial model mismatch, conventional commutation leads to large tracking error outliers, whereas robust commutation exhibits greater robustness to model discrepancies.
    \section{Experimental Results}\label{sec:experiment}
    \begin{figure}\vspace{5pt}
        \centering
        \includegraphics[width=\linewidth]{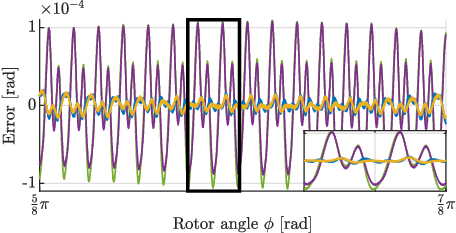}
        \caption{Experimental data of the tracking error, showing considerable tooth-by-tooth variation. When using a model $\hat{\mathbf{g}}_{\text{sine}}$, the robust function $\mathbf{f}$ (E2, \protect\purline) results in less tracking error than the conventional $\mathbf{f}_{\text{conv}}$ (E1, \protect\greenline). A similar improvement of the robust functions (E4, \protect\yelline) compared to conventional functions (E3, \protect\blueline) is seen for a model $\hat{\mathbf{g}}$ that is identified using~\cite{Meer2023b}.}\label{fig:exp_e}
        \end{figure}
        This section demonstrates experimentally that robust commutation reduces torque ripple from variation in rotor teeth.
        \subsection{Experimental setup and method}
        A single SRM is considered with $n_t=131$ teeth and $n_c=3$ coils, see~\cite{Kramer2020}. A PID controller with a bandwidth of 20 Hz is given and the sampling frequency is 10 kHz. Four constant-velocity tracking experiments are carried out with $\dot{r}=15$ teeth per second for one rotation. We use:\begin{enumerate}[label={E\arabic*:}]
            \item a conventional $\mathbf{f}_{\text{conv}}$, constructed based on a $\hat{\mathbf{g}}_{\text{sine}}$ that consists of three sinusoids, shifted 120$^{\circ}$ in phase.
            \item a robust $\mathbf{f}$, constructed for this sinusoidal model. The variance is chosen as $\boldsymbol{\Sigma}_{\theta}=5\cdot 10^{-3} \mathbf{I}$ and a model structure is a Fourier basis with 5 harmonics. 
            \item a conventional  $\mathbf{f}_{\text{conv}}$, constructed based on a $\hat{\mathbf{g}}$ which is identified using~\cite{Meer2023b}. 
            \item a robust $\mathbf{f}$, constructed for this identified model. The variance $\boldsymbol{\Sigma}_{\theta}$ thus follows from experimental data. 
        \end{enumerate}
Both robust commutation functions are created by solving~\eqref{eq:problem} using $n_\alpha=50$, $\ell=0.3$, $N=100$ and $\mu=3$. To mitigate the effect of parasitical disturbances that do not relate to commutation, e.g., bearing imperfections, we average the tracking error over 25 experiments, each with a $2\pi$ offset added to the starting position.  

\subsection{Results and analysis}
Figure~\ref{fig:exp_e} displays the error, showing significant tooth variation. Although commutation functions target the `average tooth', actual teeth differ greatly, causing torque ripple. Robust commutation reduces the median RMS error per tooth from 54.6 to 50.0 $\mu$rad for the sinusoidal model (-8\%) and from 15.4 to 13.7 $\mu$rad for the identified model (-11\%).  The reduced performance increase compared to simulations may result from unmodeled effects like sensor noise and magnetic saturation, which also induce torque ripple, thereby increasing RMS error for all commutation methods.

Combining the identification method of~\cite{Meer2023b} with this paper's robust commutation functions creates a framework that starts with a simple sinusoidal model $\hat{\mathbf{g}}_{\text{sine}}$ (E1), identifies a precise random model $\hat{\mathbf{g}}$, and then applies our commutation design method to achieve low torque ripple (E4) across all model realizations. This approach eliminates the need for torque sensors and physical modeling, providing a low-order, adaptable commutation function suitable for various SRMs of the same design without requiring individual adjustments for manufacturing differences.
\section{Conclusions and future work}\label{sec:conclusions}
A framework is developed for the offline design of robust commutation functions to achieve accurate closed-loop control of SRMs, addressing unknown variations in rotor teeth or across different SRMs. This approach uses the model variance in the torque-current-angle relationship to create a commutation function that optimally switches current between coils, reducing torque ripple due to manufacturing deviations. This reduction is significant for both large and small modeling errors, as shown by Monte Carlo simulations and experimental validation. This method is particularly suited for low-cost applications requiring a universal driver, enhancing tracking performance and reducing acoustic noise.

Future work will focus on commutation with fewer parameters, redefining the cost function and constraints on a continuous domain to remove discretization errors, and examining scenarios where the true system is outside the model class.

\section{Acknowledgements}
The authors would like to thank TNO, and in particular Lukas Kramer and Joost Peters, for the developments that have led to these results and for their help and support in carrying out the experiments reported in this paper.

\addtolength{\textheight}{-12cm}   






\bibliographystyle{IEEEtran}
\bibliography{library}

\end{document}